\documentclass[acmsmall,authorversion]{acmart}

\usepackage[utf8]{inputenc}
\usepackage{booktabs}
\usepackage{rotating}

\setcopyright{acmlicensed}
\acmJournal{CSUR}
\acmVolume{} 
\acmNumber{} 
\acmDOI{10.1145/3494522}

\copyrightyear{2021}

\newcommand{\acite}[1]{{\protect\NoHyper\citeauthor{#1}\protect\endNoHyper}~\cite{#1}}
\newcommand{\eg}{\textit{e.g.,}}
\newcommand{\ie}{\textit{i.e.,}}

\newtheorem*{def1}{Definition}

\usepackage{tikz}
\usetikzlibrary{calc,trees,positioning,shapes,shadows,arrows,fit}
\usepackage{multirow}
\usepackage{longtable}

\newcommand{\rev}[1]{\textcolor{black}{#1}}
\newcommand{\revb}[1]{\textcolor{black}{#1}}

\begin{document}

\title{A Survey on Task Assignment in Crowdsourcing}

\author{Danula Hettiachchi}
\email{danula.hettiachchi@unimelb.edu.au}
\orcid{0000-0003-3875-5727}
\affiliation{%
  \institution{The University of Melbourne}
  \city{Melbourne}
  \country{Australia}
}

\author{Vassilis Kostakos}
\email{vassilis.kostakos@unimelb.edu.au}
\orcid{0000-0003-2804-6038}
\affiliation{%
  \institution{The University of Melbourne}
  \city{Melbourne}
  \country{Australia}
}

\author{Jorge Goncalves}
\email{jorge.goncalves@unimelb.edu.au}
\orcid{0000-0002-0117-0322}
\affiliation{%
  \institution{The University of Melbourne}
  \city{Melbourne}
  \country{Australia}
}

\renewcommand{\shortauthors}{Hettiachchi et al.}


\begin{abstract}
Quality improvement methods are essential to gathering high-quality crowdsourced data, both for research and industry applications. A popular and broadly applicable  method is task assignment that dynamically adjusts crowd workflow parameters. In this survey, we review task assignment methods that address: heterogeneous task assignment, question assignment, and plurality problems in crowdsourcing. We discuss and contrast how these methods estimate worker performance, and highlight potential challenges in their implementation. Finally, we discuss future research directions for task assignment methods, and how crowdsourcing platforms and other stakeholders can benefit from them.
\end{abstract}



\maketitle

\section{Introduction}

Crowdsourcing is the process of gathering information or input of a task from a large number of individuals, typically via the Internet~\cite{Howe2006TheCrowdsourcing}. Crowdsourcing allows task requesters to access a large workforce with diverse skills and capabilities cost-effectively and efficiently compared to hiring experts or dedicated workers~\cite{Nowak2010HowCrowdsourcing}. Due to this, crowdsourcing has gained widespread popularity and has also become a critical step in harnessing training data for various machine learning models~\cite{vaughan2017making}.

As crowdsourced input originates from a multitude of workers where task requesters have limited visibility of their background information or credentials, ensuring \rev{high-quality} contributions has been \rev{a significant} research challenge. The literature proposes different quality assurance methods, including training workers~\cite{doroudi2016toward}, providing feedback~\cite{Dow2012,Goncalves2013IncluCity:Accessibility}, improving task design~\cite{Dimara2017NarrativesinCrowdsourcedEvaluation,Gadiraju2017ClarityCrowdsourcing}, implementing task workflows~\cite{Little2010ExploringProcesses}, aggregating responses~\cite{Zheng2017TruthInference,Goncalves2016CrowdsourcingSitu}, and detecting outliers~\cite{Jung2011ImprovingVoting,Hosio2018CrowdsourcingPain}. Among such methods, matching workers with compatible tasks or `task assignment' has emerged as an important mechanism that can increase the quality of the contributed data~\cite{Kittur2013TheWork}. 

While there exist several other surveys related to data quality in crowdsourcing~\cite{Zhang2016,Daniel2018,Li2016CrowdsourcedSurvey,vaughan2017making}, none of them extensively review assignment methods. Our review provides an overview of data quality improvement approaches in crowdsourcing, organised under pre-execution, online and post-processing methods. Then, we dive deep into task assignment approaches. Particularly, we discuss different methods of modelling and estimating worker performance that takes place prior to the task assignment step. We also distinguish question assignment (match individual questions within the task based on factors like question difficulty, worker quality and current answer confidence) from heterogeneous task assignment where we match workers to specific types of tasks (\eg~ image classification, sentiment analysis). 

Optimum task assignment is a challenging endeavour due to variations in crowd tasks, the inconsistencies with the availability and the diversity of the worker population. Therefore, researchers present various methods that utilise \rev{historic worker data}~\cite{Mo2013CrosstaskCrowdsourcing}, current answer distribution~\cite{Fan2015,Khan2017CrowdDQS}, gold standard questions (questions with known answers)~\cite{Ipeirotis2014quizz}, worker attributes~\cite{Kazai2011,Goncalves2017TaskRouting}, and behavioural data~\cite{Han2016,Rzeszotarski2011}. Our review sheds light on how these different methods perform under different scenarios. Furthermore, we discuss broader challenges and limitations of assignment approaches, and present future directions for research on task assignment.

Overall, our survey makes the following contributions:
\begin{itemize}
    \item We provide a detailed overview on existing crowdsourcing data quality improvement techniques that aim to match workers with compatible tasks and questions.
    \item We identify and review specific methods that address task assignment, question assignment, and plurality problems.
    \item We discuss challenges in employing different worker performance estimation and assignment methods in a crowdsourcing platform.
\end{itemize}

\subsection{Outline of the Survey}

Section~\ref{sec:literature-selection} describes the method \rev{followed in selecting} the literature included in this survey. Section~\ref{sec:quality-enhancement} briefly reviews data quality improvement methods in crowdsourcing, and Section~\ref{sec:problems} defines the four task assignment problems that we discuss in the survey. Section~\ref{sec:worker-performance-estimation} elaborates on worker performance modelling and estimation methods, which are two critical steps of task assignment. Then, Section~\ref{sec:task-assignment} summarises task assignment approaches, including heterogeneous task assignment, question assignment, the plurality problem and budget allocation methods. \rev{Section~\ref{sec:platforms} provides an overview of existing crowdsourcing platforms and their available task assignment methods.} Finally, Sections~\ref{sec:future}~\&~\ref{sec:conclusion} provide future directions on data quality research in crowdsourcing and concluding remarks of our survey.

\section{Literature Selection}
\label{sec:literature-selection}
\subsection{\rev{Background and Motivation}}

We note several related surveys that capture different elements of crowdsourcing. \acite{Daniel2018} look at overarching quality enhancement mechanisms in crowdsourcing. Their survey organises literature under three segments: quality model, which describes different quality dimensions, quality assessment methods, and quality assurance actions. While \acite{Daniel2018} summarise task assignment methods, they are not analysed in detail due to the broader scope of their survey.

\acite{Zheng2017TruthInference} examine 17 truth inference techniques such as majority vote, Zencrowd~\cite{Demartini2012ZenCrowd} and Minimax~\cite{Zhou2012LearningEntropy}. The survey also presents an evaluation of different methods using five real work datasets. The primary focus of our survey lies outside truth inference methods. However, we provide a summary of truth inference methods in Section~\ref{sec:post-processing}, under post-processing data quality improvement methods.

\acite{Li2016CrowdsourcedSurvey} surveys crowdsourced data management with an emphasis on different crowd data manipulation operations such as selection, collection and join. Their survey organises prior work under quality, cost and latency control methods. \acite{vaughan2017making} also present a comprehensive review on how crowdsourcing methods can benefit machine learning research. 

Overall, in contrast to prior literature reviews, our survey sheds light on the task assignment problem in crowdsourcing and discusses related assignment based quality improvement methods. \rev{In particular, our survey examines the research questions outlined in Table~\ref{tab:research-questions}.}

\begin{table}[h]
    \centering
    \caption{\rev{Research questions examined in the survey}.}
    \label{tab:research-questions}
\begin{tabular}{p{4.8cm}p{8.5cm}}
\toprule
\rev{Research Question} & \rev{Description}\\
\midrule
\rev{What are the different ways of matching workers with tasks in crowdsourcing and specific methods proposed to achieve them?} &
\rev{Our survey differentiates task assignment from generic data quality improvement, identifies and defines types of task assignment problems, and provides a detailed review of proposed approaches.}\\
\midrule
\rev{How do we estimate and model worker performance for task assignment?} &
\rev{Performance estimation is an essential step in crowdsourcing task assignment. By dissecting proposed methods into estimation and assignment steps, we provide a detailed outlook of different estimation and modelling methods that can also promote the reuse of components in future implementations.}\\
\midrule
\rev{What are the challenges and limitations of task assignment methods and their availability in existing crowdsourcing platforms?} & 
\rev{While many data quality improvement methods have been proposed in the literature, not many of them have been widely adopted in commercial crowdsourcing platforms. We review factors that limit their practical uptake and detail specific task assignment methods available in these platforms.}\\
\bottomrule

\end{tabular}
\end{table}

\subsection{Literature Selection}

We conducted an extensive literature search on the ACM Digital Library using a query that includes keywords `task assignment', `task routing' or `data quality' and `crowd*' in the Abstract. We included articles published from 2010 and retrieved 747 records. We reduced the resulting set of papers by limiting to publications from a list of \rev{conferences} and journals that, to the best of our knowledge, publish work on crowdsourcing and related topics. Selected conferences were AAAI, AAMAS, CHI, CIKM, CSCW, ESEM, HCOMP, HT, ICDE, ICML, IUI, JCDL, KDD, SIGIR, SIGMOD, UbiComp, UIST, WI, WSDM and WWW. Selected journals were PACM IMWUT, PACM HCI, TKDE, TSC, VLDB. We also excluded workshops, demo papers, posters, extended abstracts, etc. Literature from specific venues that are not included in the ACM Digital Library (\eg~HCOMP) \rev{was} manually screened and added to our dataset. Then, we carefully inspected the remaining papers and filtered out papers that were deemed to not be relevant. Furthermore, the survey also includes several additional papers hand-picked by the authors due to their relevance to the topic.

\subsection{Scope}

Crowdsourcing extends beyond traditional online crowdsourcing using desktop or laptop computers. Other general types which can overlap include mobile crowdsourcing~\cite{musthag2013labor} (\eg~smartphones, tablets), situated crowdsourcing~\cite{Goncalves2013CrowdsourcingontheSpot,Hosio2014SituatedModel,Goncalves2014ProjectiveEmotion} (\eg~public displays), spatial crowdsourcing~\cite{Tong2020SpatialSurvey,Goncalves2017CrowdPickUp} (\eg~workers attempt \rev{location-based} tasks including physical tasks) \rev{and crowdsensing~\cite{Guo2015MobileCrowdSensing} (\eg~ workers passively contribute sensor data from mobile devices)}. Task assignment in crowdsourcing has also been investigated based on such domains. However, due to wide variations in techniques used in these different settings, we limit our scope to online crowdsourcing.

Crowdsourcing can also be broadly categorised as paid and unpaid crowd work based on the rewards received by workers. Paid work corresponds to crowdsourcing tasks where workers receive monetary rewards typically through a crowdsourcing platform \rev{that} facilitates the payment process. Unpaid or voluntary crowd work is also completed in popular platforms and projects like Wikipedia\footnote{https://www.wikipedia.org}, Moral Machine~\cite{Awad2020CrowdsourcingMachines}, \rev{Crowd4U~\cite{Ikeda2016CollaborativeCrowd4U}, Zooniverse\footnote{https://www.zooniverse.org},} and Test My Brain~\cite{Germine2012IsExperiments}. However, there are key distinctions in how you motivate unpaid and paid crowd work~\cite{Mao2013VolunteeringCrowdsourcing,Rogstadius2011,Goncalves2015MotivatingCrowdsourcing}. For example, in Test My Brain, workers get personalised feedback that \rev{helps} them learn more about their mind and brain. In this review, we primarily focus on methods and literature that investigate paid crowdsourcing tasks on commercial crowdsourcing platforms.

When we consider the type of work available on crowdsourcing platforms, they can range from micro tasks~\cite{difallah2015dynamics} such as labelling, ranking and classification to complex and long term tasks like software and web development tasks~\cite{Stol2014TwosDevelopment}. Our survey focuses on crowdsourcing techniques concerning tasks that can be completed in a single session, which constitutes the bulk of available crowd work.

\section{Quality Enhancement in Crowdsourcing}
\label{sec:quality-enhancement}

As crowdsourcing typically relies on contributions from a diverse workforce where task requesters have limited information on the workers, it is important to employ data quality improvement measures~\cite{Daniel2018}. In this section, we provide an overview of data quality in crowdsourcing. 

In crowdsourcing, data quality is typically quantified via different attributes such as task accuracy, the response time of collected data, and cost-efficiency. Different quality improvement methods aim to improve one or more quality attributes. For example, the accuracy of a translation task can be enhanced in a cost-effective manner by employing workflow changes~\cite{Ambati2012Collaborative}.

We note that quality improvement methods can differ from one another based on the following characteristics.

\begin{itemize}
    \item \textit{Applicability:} A quality improvement method can work for a specific type of task, a broader range of tasks or across all types of tasks. Universal methods are highly desired, yet can be costly and difficult to implement. For example, certain question assignment methods~\cite{Ipeirotis2014quizz,Fan2015} only work for multi-class labelling tasks. In contrast, worker filtering based on approval rate works for most tasks when worker-history is available. 
    \item \textit{Complexity:} Some quality improvement methods involve complex implementations that require substantial time and effort. Such methods are not suitable for one-time jobs. For example, it is not straightforward to implement crowd workflows that facilitate real-time discussions among workers~\cite{Chen2019Cicero:Crowdsourcing,Hosio2015CrowdsourcingOulu}.
    \item \textit{Effectiveness:} The effectiveness of quality improvement methods also varies. The effectiveness of a method can be quantified by measuring the quality attributes.
    \item \textit{Cost:} There is an inherent cost attached to each quality improvement method. It is explicit for some methods (\eg~issuing bonus payments to workers), while others have indirect costs (\eg~infrastructure cost to capture and analyse worker behaviour data).
\end{itemize}

Generally, task requesters prefer quality improvement methods that are low in complexity, highly effective, economical and broadly applicable. However, methods that satisfy all these quality needs are scarce, and task requesters typically select quality improvement methods based on the specific task at hand, time and budget constraints, quality requirement and platform compatibility.

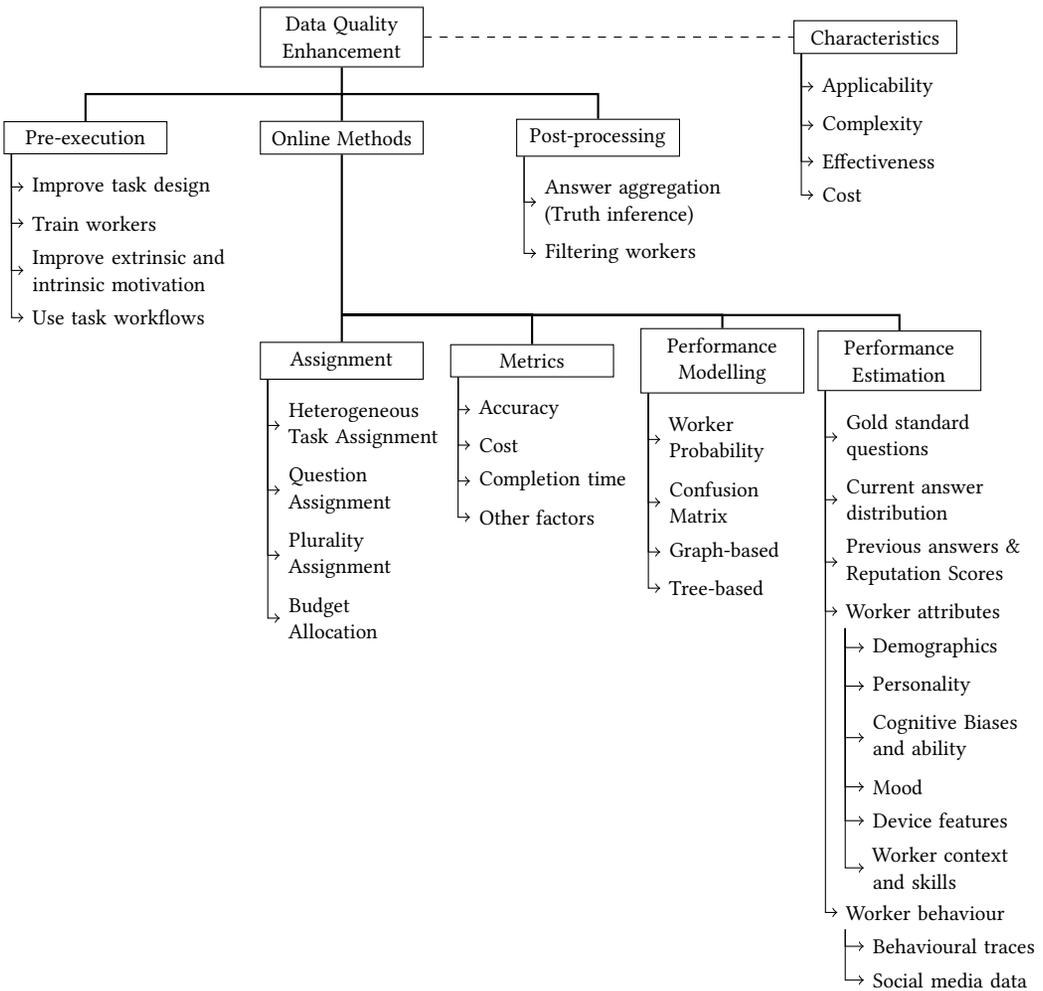
\begin{figure}[b]
\centering
\begin{tikzpicture}
\tikzset{every node/.style={align=center, minimum height=10pt, text width=55pt, font=\footnotesize}}
\tikzset{leaf node/.style={align=left, text width=80pt, font=\footnotesize}}

\node[,draw=black] (b1) {Pre-execution};
\node[right=35pt,draw=black] (b2) at (b1.east) {Online Methods};
\node [leaf node, below=5pt, xshift=20pt] (b11) at (b1.south) {Improve task design};
\node [leaf node, below=0pt] (b12) at (b11.south) {Train workers};
\node [leaf node, below=0pt] (b13) at (b12.south) {Improve extrinsic and\\intrinsic motivation};
\node [leaf node, below=0pt] (b14) at (b13.south) {Use task workflows};

\node[right=35pt,draw=black] (b3) at (b2.east) {Post-processing}; 
\node [leaf node, below=5pt, xshift=20pt] (b31) at (b3.south) {Answer aggregation\\(Truth inference)};
\node [leaf node, below=0pt] (b32) at (b31.south) {Filtering workers};

\node [below=70pt,draw=black] (c1) at (b2.south) {Assignment};
\node [leaf node, below=5pt, xshift=20pt] (d11) at (c1.south) {Heterogeneous\\Task Assignment};
\node [leaf node, below=0pt] (d12) at (d11.south) {Question\\ Assignment};
\node [leaf node, below=0pt] (d13) at (d12.south) {Plurality\\ Assignment};
\node [leaf node, below=0pt] (d14) at (d13.south) {Budget\\ Allocation};

\node [right=10pt,draw=black] (c2) at (c1.east) {Metrics};
\node [leaf node, below=5pt, xshift=20pt] (d21) at (c2.south) {Accuracy};
\node [leaf node, below=0pt] (d22) at (d21.south) {Cost};
\node [leaf node, below=0pt] (d23) at (d22.south) {Completion time};
\node [leaf node, below=0pt] (d24) at (d23.south) {Other factors};

\node [right=10pt,draw=black] (c3) at (c2.east) {Performance Modelling};
\node [leaf node, below=5pt, xshift=20pt] (d31) at (c3.south) {Worker\\Probability};
\node [leaf node, below=0pt] (d32) at (d31.south) {Confusion\\Matrix};
\node [leaf node, below=0pt] (d33) at (d32.south) {Graph-based};
\node [leaf node, below=0pt] (d34) at (d33.south) {Tree-based};

\node [right=5pt,draw=black] (c4) at (c3.east) {Performance Estimation};
\node [leaf node, below=5pt, xshift=20pt] (d41) at (c4.south) {Gold standard\\questions};
\node [leaf node, below=0pt] (d42) at (d41.south) {Current answer\\distribution};
\node [leaf node, below=0pt] (d45) at (d42.south) {Previous answers \&\\Reputation Scores};
\node [leaf node, below=0pt] (d43) at (d45.south) {Worker attributes};
\node [leaf node, below=0pt, xshift=10pt,minimum height=10pt] (d431) at (d43.south) {Demographics};
\node [leaf node, below=0pt,minimum height=10pt] (d432) at (d431.south) {Personality};
\node [leaf node, below=0pt,minimum height=10pt] (d433) at (d432.south) {Cognitive Biases \\and ability};
\node [leaf node, below=0pt,minimum height=10pt] (d434) at (d433.south) {Mood};
\node [leaf node, below=0pt,minimum height=10pt] (d435) at (d434.south) {Device features};
\node [leaf node, below=0pt,minimum height=10pt] (d436) at (d435.south) {Worker context \\and skills};

\node [leaf node, below=100pt] (d44) at (d43.south) {Worker behaviour};
\node [leaf node, below=0pt, xshift=10pt,minimum height=10pt] (d441) at (d44.south) {Behavioural traces};
\node [leaf node, below=0pt,minimum height=10pt] (d442) at (d441.south) {Social media data};

\node[above=20pt,draw=black] (top) at ($(b2.north)$) {Data Quality \\Enhancement};

\node[right=140pt,draw=black] (a2) at (top.east) {Characteristics};
\node [leaf node, below=5pt, xshift=20pt] (a21) at (a2.south) {Applicability};
\node [leaf node, below=0pt] (a22) at (a21.south) {Complexity};
\node [leaf node, below=0pt] (a23) at (a22.south) {Effectiveness};
\node [leaf node, below=0pt] (a24) at (a23.south) {Cost};

\coordinate (atop) at ($(top.south) + (0,-10pt)$);
\coordinate (btop) at ($(b3.south) + (0,-10pt)$);
\coordinate (c1top) at ($(c1.north) + (0,10pt)$);

\draw[thick] (top.south) -- (atop)
(b1.north) |- (atop)
(b2.north) |- (atop) -| (b3.north)
(b2.south) -- (c1.north)
(c2.north) |- (c1top)
(c3.north) |- (c1top)
(c4.north) |- (c1top);

\draw[dashed] (top.east) -- (a2.west);


\foreach \val in {1,2,3,4}
\draw[->] ($(a2.south west) + (3pt,0)$) |- (a2\val.west);

\foreach \val in {1,2,3,4}
\draw[->] ($(b1.south west) + (3pt,0)$) |- (b1\val.west);

\foreach \val in {1,2}
\draw[->] ($(b3.south west) + (3pt,0)$) |- (b3\val.west);

\foreach \hval in {1,2,3,4}
\foreach \val in {1,2,3,4}
\draw[->] ($(c\hval.south west) + (3pt,0)$) |- (d\hval\val.west);

\draw[->] ($(c4.south west) + (3pt,0)$) |- (d45.west);

\foreach \val in {1,2,3,4,5,6}
\draw[->] ($(d43.south west) + (3pt,0)$) |- (d43\val.west);

\foreach \val in {1,2}
\draw[->] ($(d44.south west) + (3pt,0)$) |- (d44\val.west);

\end{tikzpicture}
\caption{\rev{Overview of quality enhancement methods and related concepts discussed in the survey.}}
\label{fig:overview}
\end{figure}

While there is a wide array of such quality enhancement techniques, based on the method execution phase, they can be broadly categorised into pre-execution methods, online methods and post-processing techniques as detailed in \rev{Figure~\ref{fig:overview}}. Given the standard crowdsourcing workflow, task requesters consider and employ pre-execution methods before task deployment. Fundamentally, through these methods, requesters specify how the task should be presented and executed in the crowdsourcing platform. Next, online methods alter the crowd task execution by dynamically deciding parameters such as the number of labels to collect, worker-task assignment, and task reward. Finally, post-processing methods examine how we can obtain better outcomes by processing the gathered crowd input. In this survey, we are primarily interested in online methods, however we briefly summarise pre-execution and post-processing methods in the following sub-sections. \rev{In addition, Figure~\ref{fig:overview} provides an overview of different concepts and categories related to online assignment.}

\subsection{Pre-execution Methods}

Data quality improvement methods employed at the pre-execution phase involve improving how workers interact with the task in terms of task design and crowdsourcing workflows. 

\subsubsection{Task Design and Crowdsourcing Workflows}

Improving task design based on design guidelines and crowdsourcing best practices is one of the most well-known quality improvement methods. 
Research shows that clear task descriptions~\cite{Gadiraju2017ClarityCrowdsourcing}, data semantics or narratives that provide task context~\cite{Dimara2017NarrativesinCrowdsourcedEvaluation}, and enhanced task user interfaces that improve  usability~\cite{aCampo2019CommunityPlatforms,Alonso2011DesignCrowdsourcing} and reduce cognitive load~\cite{AlagaraiSampath2014CognitivelyInspiredTask} elevate data quality. 

The outcomes of methods relating to task design can vary depending on the task itself. For example, Find-Fix-Verify~\cite{Bernstein2010Soylent} is a workflow introduced for writing tasks such as proofreading, formatting and shortening text. Iterate and vote is another design pattern where we ask multiple workers to work on the same task in a sequential manner. \acite{Little2010ExploringProcesses} shows that the iterate and vote method works well on brainstorming and transcription tasks. Similarly, under map-reduce, a larger task can be broken down into discrete sub-tasks and processed by one or more workers. The final outcome is obtained by merging individual responses~\cite{Kittur2011CrowdForge:Work,Cheng2015BreakItDown}.

Many other complex workflows have been proposed. For instance, the assess, justify \& reconsider~\cite{Drapeau2016MicroTalk:Accuracy} workflow improves task accuracy by 20\% over the majority vote for annotation tasks. Several extensions to this method have been proposed, such as introducing multiple turns~\cite{Schaekermann2018ResolvableWork,Chen2019Cicero:Crowdsourcing}. Annotate and verify is another workflow that includes a verification step. \acite{Su2012CrowdsourcingDetection} show that data quality in a bounding box task is improved when they employ the annotate and verity method with two quality and coverage assessment tasks followed by the drawing task~\cite{Su2012CrowdsourcingDetection}.

More complex workflows that facilitate real-time group coordination \cite{Schaekermann2018ResolvableWork,Chen2019Cicero:Crowdsourcing} can be challenging to incorporate into a crowdsourcing platform. Other variants include tools that allow workers~\cite{Kulkarni2012} and task requesters (\eg~Retool~\cite{Chen2017ReTool:Demonstration}, CrowdWeaver~\cite{Kittur2012CrowdWeaver:Work}) to design custom workflows. There is limited work that explores how to build and manage the crowdsourcing pipeline when employing a task workflow~\cite{Tran-Thanh2015CrowdsourcingConstraints}. For example, the reward for each step can be dynamically adjusted to efficiently process the overall pipeline~\cite{Mizusawa2018EfficientWorkflows}. On the contrary, some work argues that static crowdsourcing workflows are limited in terms of supporting complex work and calls for open-ended workflow adaptation~\cite{Retelny2017NoEnough}.

Other related task design and workflow improvements include gamification~\cite{Morschheuser2016GamificationReview,Goncalves2014GameDisplays} and adding breaks or micro-diversions~\cite{Dai2015MicroDiversions}.

\subsubsection{Feedback and Training}

Providing feedback to workers based on their work can improve the data quality in crowdsourcing. \acite{Dow2012} report that external expert feedback and self-assessment encourages workers to revise their work. 
\acite{Dow2012} highlight three key aspects of feedback for crowd work. 
`Timeliness' indicates when the worker gets feedback (\ie~synchronously or asynchronously). The level of detail in the feedback or `specificity' can vary from a simple label (\eg~approve, reject) to more complex template-based or detailed one to one feedback.
Finally, `source' or the party giving feedback, which can be experts, peer workers, the requester, or the worker themselves.

In a peer-review setup, the process of reviewing others' work has also been shown to help workers elevate their own data quality~\cite{Zhu2014}. Similarly, workers achieve high output quality when they receive feedback from peers in an organised work group setting~\cite{whiting2017crowd}. While expert and peer feedback are effective in improving data quality, it is challenging to ensure the timeliness of feedback which is important when implementing a scalable feedback system.

It is also possible to deploy a feedback-driven dedicated training task and let workers complete multiple training questions until they achieve a specified data quality threshold. ~\acite{park2014toward} report that such a mechanism can be effective in crowdsourcing tasks that involve complex tools and interfaces. However, training or feedback may also bias the task outcome depending on the specific examples selected for the training/feedback step ~\cite{Le2010EnsuringDistribution}. Feedback can also be used to explain unclear task instructions. For example, prior work by \acite{manam2018wingit} proposes a Q\&A and Edit feature that workers can use to clarify and improve task instructions or questions.

Other similar work tools that can potentially help improve data quality include third-party web platforms, browser extensions and scripts (\eg~Turkopticon~\cite{Irani2013Turkopticon:Turk}, Panda Crazy\footnote{https://github.com/JohnnyRS/PandaCrazy-Max})~\cite{Kaplan2018StrivingWorkers}. These tools provide additional information for workers to avoid substandard tasks and make their work more efficient.

\subsection{Online Methods}

While pre-execution methods focus on priming the task and workers, online methods aim to increase data quality by dynamically changing task deployment parameters and conditions like matching workers with compatible and relevant tasks. In this survey, we primarily focus on such online assignment methods, that we discuss in detail in the Sections~\ref{sec:problems},~\ref{sec:worker-performance-estimation}~\&~\ref{sec:task-assignment}.

\subsection{Post-processing Methods}
\label{sec:post-processing}

Post-processing methods are employed after workers complete the entire batch of tasks in the crowdsourcing platform. A large portion of post-processing methods falls under answer aggregation techniques. We also discuss several other methods, including filtering workers. 

\subsubsection{Aggregating Answers}

Typically in crowdsourcing, we obtain multiple answers for each question. Once all the answers are collected, we need to aggregate them to create the final answer for each question. This process is also known as truth inference in crowdsourcing. There are many ways to aggregate answers, and task requesters may opt for different strategies depending on the task and data quality needs.

Majority voting is the most simple and naive, yet widely used approach for answer aggregation~\cite{Zheng2017TruthInference}. However, majority vote can fail when only a handful of highly accurate workers provide the correct answer. Prior work has proposed many extensions to majority voting. For example, instead of calculating the majority vote, the labels can be aggregated to a score that reflects the level of agreement~\cite{Zheng2017TruthInference}. Then, we can calculate the best threshold value to obtain the final answer. A training set or a gold standard question set can be used when determining the threshold.

\acite{Zhuang2015} examined the bias that can be introduced into crowdsourcing when a worker provides answers to multiple tasks grouped into a batch, which is a common mechanism employed to reduce cost and improve convenience for the worker. They proposed an alternative to majority voting, which could result in improved accuracy when batching is present. \acite{Ma2015FaitCrowd} proposed a truth inference method that is able to account for the varying expertise of workers across different topics. 

For rating and filtering tasks, \acite{DasSarma2016} proposed an algorithm for finding the global optimal estimates of accurate task answers and worker quality for the underlying maximum likelihood problem. They claim their approach outperforms Expectation Maximisation based algorithms when the worker pool is sufficiently large. Further, in an extensive survey on truth inference, \acite{Zheng2017TruthInference} evaluate the performance of different truth inference algorithms.

\subsubsection{Clustering}

\acite{Kairam2016PartingCrowds} proposed an automated clustering-based method as a design pattern for analysing crowd task responses. Using entity annotations of Twitter posts and Wikipedia documents, they identified systematic areas of disagreement between groups of workers that can be used to identify themes and summarise the responses. 

\subsubsection{Filtering Answers}

After data collection, we can also remove specific responses to improve the data quality.
For example, If we are able to identify malicious workers who may submit purposely inaccurate or incomplete responses, we can filter all the answers provided by such users during the aggregation process. Instead of using worker responses as the sole quality signal, \acite{Moshfeghi2016IdentifyingPlatforms} propose a method that uses task completion time to \rev{identify} careless workers. Similarly, post-hoc worker filtering is also possible after estimating worker accuracy through different techniques, such as analysing worker behavioural traces~\cite{Rzeszotarski2011,Han2016} and the worker network~\cite{Kuang2020APlatforms}. In Section~\ref{sec:worker-performance-estimation-sub}, we discuss estimation methods in detail. Furthermore, data quality can be impacted when workers use bots to provide automated responses or collude with other workers to share information~\cite{Checco2020AdversarialControl,Difallah2012MechanicalPlatforms}. \acite{KhudaBukhsh2014DetectingCrowdsourcing} propose an unsupervised collusion detection algorithm that can help identify such workers and remove corresponding responses. It is also possible to detect colluding \rev{workers} by analysing contribution similarity~\cite{Kamhoua2018ApproachCrowdsourcing}.  
In addition, sybils or bots can be identified by estimating worker similarity and clustering them into groups~\cite{Yuan2017SybilPlatforms}.

\section{Task Assignment Problems}
\label{sec:problems}

\begin{figure}[h]
\centering
\begin{tikzpicture}

\tikzset{every node/.style={align=center, minimum height=10pt, text width=55pt, font=\footnotesize}}

\node [] (q11) {Task ($t_1$)};
\node [below=0pt, draw] (q12) at (q11.south) {Question ($q_1$)};
\node [below=5pt, draw] (q13) at (q12.south) {Question ($q_2$)};
\node[inner sep=5pt, fit=(q11)(q12)(q13), draw](t1) {};

\node [right=20pt] (q21) at (q11.east){Task ($t_2$)};
\node [below=0pt, draw] (q22) at (q21.south) {Question};
\node [below=5pt, draw] (q23) at (q22.south) {Question};
\node[inner sep=5pt, fit=(q21)(q22)(q23), draw](t2) {};

\node [right=50pt, draw] (r1) at ($(q21.east) + (0,5pt)$) {Requester};
\node [left=50pt, draw] (w1) at (q11.west) {Worker ($w_1$)};

\node [below=25pt] (a1) at (r1.south){Answers};
\node [below=0pt, text width=70pt, draw] (a2) at (a1.south) {Answer ($A_{q1,w1}$)};
\node [below=5pt,text width=70pt, draw] (a3) at (a2.south) {Answer ($A_{q2,w1}$)};
\node[inner sep=5pt, fit=(a1)(a2)(a3), draw](ans) {};

\node[inner sep=16pt,fit=(t1)(t2), draw, label={[text width=100pt]Crowdsourcing Platform}](platform) {};

\coordinate (t1top) at ($(t1.north) + (0,5pt)$);

\draw[dashed] (w1.south) |- node[near end,below,text width=100pt]{\footnotesize Provide answers} (q12.west)
(w1.north) |- node[near end,above,text width=100pt]{\footnotesize Receive rewards} ($(platform.west) + (0,30pt)$)
(t1.north) -- (t1top)
(t1.south) |- (ans.west)
(r1.south) -- node[text width=100pt]{\footnotesize Obtain answers} (ans.north)
(t1top) -| node[near start,above,text width=100pt]{\footnotesize Create the task} (r1.north);


\end{tikzpicture}
\caption{\rev{Components of a standard crowdsourcing workflow and the relationships among them.}}
\label{fig:crowd-workflow}
\end{figure}
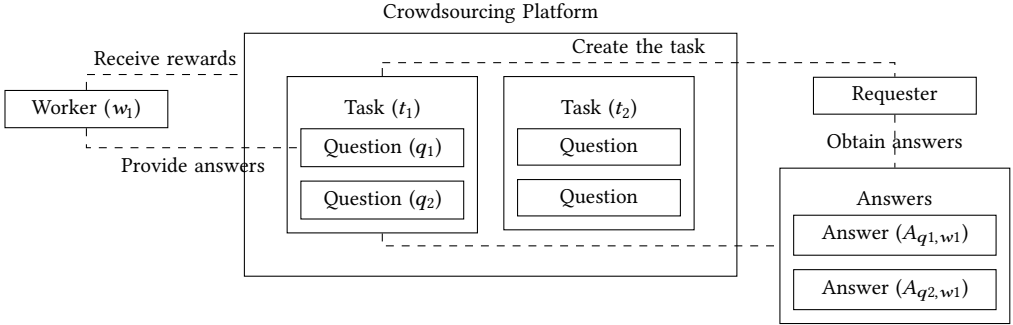

Before we examine online methods in detail, it is important to identify the different stakeholders and parameters involved. We explain the crowdsourcing workflow \rev{(Figure~\ref{fig:crowd-workflow})}, involved entities and different parameters that can be optimised in an online setting for task assignment purposes.

\begin{itemize}
    \item \textit{Requester:} A person who posts tasks on a crowdsourcing platform. Requesters reward the workers through the platform when they provide answers to their task.
    \item \textit{Worker:} A person who completes tasks on a crowdsourcing platform in return for a reward. There is a large body of literature that examines characteristics of worker population~\cite{Difallah2018Demographics,Ross2010WhoTurk}, work practices~\cite{Williams2019TheCrowdwork} and challenges faced by workers~\cite{salehi2015wedynamo}.
    \item \textit{Task:} A collection of questions of the same task type. Prior work~\cite{Gadiraju2014AWeb} has identified different task categories, such as verification and validation, interpretation and analysis, content creation, surveys, and content access. 
    \item \textit{Question:} An individual question within a task. For example, in an Audio Annotation task, this would be an audio clip that requires an annotation. An arbitrary number of answers can be collected for each question. Typically this threshold or the number of labels required for each question is pre-determined by the requester.
    \item \textit{Answer:} The answer provided by a specific worker to a specific question. Answer could take different forms depending on the task (\eg~a label `Positive' in sentiment analysis). Typically in crowdsourcing, multiple workers provide answers for the same question. Numerous metrics such as accuracy, response time can be used to measure the quality of an answer.
    \item \textit{Reward:} There can be intrinsic and extrinsic rewards~\cite{Rogstadius2011}. The main reward mechanism used in crowdsourcing includes providing a pre-specified base payment and bonus payments issued at requesters discretion. 
    \item \textit{Crowdsourcing Platform:} Interaction between workers and task requesters is often managed by a third-party platform. For example, Amazon Mechanical Turk, Appen, Prolific and Toloka are commercial crowdsourcing platforms, that charge a fee from task requesters for managing the crowdsourcing workflow.
\end{itemize}

As detailed in Table~\ref{tab:notations}, we use a consistent notation throughout the survey to describe different assignment problems.

\begin{table}[h]
  \caption{Notations used in the survey}
  \label{tab:notations}
  \begin{tabular}{ll}
    \toprule
    Set of workers & $W=\{w_1,..,w_n\}$\\
    Set of tasks & $T=\{t_1,..,t_n\}$ \\
    Set of questions for task $t$ & $Q_t=\{q_1,..,q_n\}$ \\
    A task assignment & $\{t,w\}$ \\
    A question assignment of question $q$ and worker $w$ & $QA_{q,w}$ \\
    An answer provided by worker $w$ to question $q$ & $A_{q,w}$\\
    Reward or payment for a question $q$ & $R_q$\\
  \bottomrule
\end{tabular}
\end{table}

While the interaction between entities detailed above can vary depending on the specific crowdsourcing platform, next we summarise a typical crowdsourcing workflow. Task requesters first post their tasks in a crowdsourcing platform, with specific instructions and rewards for successful completion. Workers who have already signed up in the platform can browse and start working on tasks that they are eligible for. Eligibility constraints (\eg~ location, skill and quality requirements) are often set by requesters or the crowdsourcing platform itself. Finally, when the work is completed, requesters can obtain the worker input or data contributions from the platform and compute the final output. Optionally, they may indicate whether individual worker answers meet their expectation. For instance, requesters can 'approve' or 'reject' answers. The crowdsourcing platform then transfers the reward to workers. This is similar to a first-come-first-serve or a market model. Online assignment methods in crowdsourcing aim to alter this market model by directing workers to relevant and compatible tasks in order to increase the overall data quality. At a high level, we identify and examine four key assignment challenges; heterogeneous task assignment, question assignment, plurality assignment problem and budget allocation. 

\subsection{Heterogeneous Task Assignment Problem}

The aim of heterogeneous task assignment or simply `task assignment' is to select the best-suited task for a worker when there are different tasks available (\eg~Sentiment Analysis, Entity Resolution, and Classification).

\begin{def1}
Assume that we have a set of tasks $T=\{t_1,..,t_k\}$ and a set of workers $W=\{w_1,..,w_m\}$ where $|T|=k$ and $|W|=m$. Each task $t$ may contain an arbitrary number of questions. In order to maximise the overall quality of the data we gather, for each worker $w \in W$, we aim to assign the task $t'$ where the worker is more likely to produce results of better quality. 
\end{def1}

\subsection{Question Assignment Problem}

Select a specific number of questions from a task for a worker. For example, in a Twitter Sentiment Analysis task with 1000 tweets, the aim is to find specific tweets to assign for each worker.

\begin{def1}
Assume that we have a set of questions $Q=\{q_1,..,q_k\}$ for a specific task $t$ and a set of workers $W=\{w_1,..,w_m\}$ where $|Q|=k$ and $|W|= m$. In order to maximise the overall quality of the data we gather, for each worker, we aim to assign one or several questions where the worker is more likely to produce results of better quality.
\end{def1}

\subsection{Plurality Assignment Problem}

Deciding on the optimal number of workers that should be assigned to each sub-task or question is known as plurality assignment problem. Typically in crowdsourcing platforms, requesters manually configure a fixed number as the number of workers to be assigned for each task.

\begin{def1}
Assume that we have a set of questions $Q=\{q_1,..,q_k\}$ for a specific task $t$ and a set of workers $W=\{w_1,..,w_m\}$ where $|Q|=k$ and $|W|= m$. For each question $q \in Q$, multiple workers can provide answers (\eg~$A_{q,w1}$, $A_{q,w2}$, .. $A_{q,wx}$). We want to determine the ideal number of answers needed for each question $q$.
\end{def1}

\subsection{Budget Allocation}

The wide popularity of crowdsourcing is largely due to its economical nature when compared to other ways to acquiring large volumes of data. Hence, in addition to the data quality, budget allocation is an important factor in crowd work. Certain work considers budget allocation as part of the task or question assignment problem. For example, \acite{Assadi2015OnlineMarkets} investigate task assignment with the aim of maximising the number of tasks allocated with a fixed budget.
\section{Worker Performance Estimation}
\label{sec:worker-performance-estimation}

Worker performance estimation is a critical step in online assignment process. If performance estimations are unreliable, subsequent task, question or budget assignment decisions will not lead to desired quality enhancements. In this section, we discuss different metrics that can be used for estimation, data structures utilised for worker performance modelling and ways of estimating the performance.

\subsection{Performance Metrics}

\subsubsection{Accuracy}

Task accuracy is the most widely used performance metric in crowdsourcing. Accuracy is typically a number between 0 (incorrect) and 1 (correct) and can be defined in different ways depending on the task. For instance, for a classification task with single correct answer, accuracy of each question would be 1 if the worker provides the correct label and 0 otherwise. In contrast, a distant metric can define the similarity between text for translation tasks which results in a fraction.
Other metrics that represent task accuracy include F-score~\cite{Zheng2015}, information gain~\cite{Li2014} for multiple-choice tasks, mean Intersection over Union (mIoU) for image annotation tasks~\cite{papadopoulos2017extreme}, etc.

\subsubsection{Cost}

While there are different crowd pricing mechanisms discussed in the literature~\cite{Singer2013PricingMechanisms}, in a typical crowdsourcing platform, there is a pre-specified cost attached to each collected answer. However, other costs such as bonus payments, platform fees (\eg~MTurk\footnote{https://www.mturk.com/pricing}) can increase the total cost. Since crowdsourcing is often used for tasks with a large number of questions, cost is considered an important performance metric. 

\subsubsection{Task Completion Time}

When we consider task completion, there are two key metrics, time that workers spend on completing each question (\ie~work time) and total time needed to complete a task job that contains a set of questions (\ie~batch completion time). Both metrics can be optimised in different ways. Minimising work time is particularly helpful for tasks that require workers with specific skills or backgrounds~\cite{Mavridis2016UsingCrowdsourcing}. In addition to task assignment, task scheduling strategies also aim to optimise batch completion time~\cite{Difallah2019Deadline-AwareSystems}. Crowdsourcing platforms typically provide task time information to requesters and they can also set a maximum time limit for each question. 

\subsubsection{Other Factors}

Another indirect performance metric is worker satisfaction. Prior work highlights a relationship between crowd worker satisfaction and turnover~\cite{Brawley2016WorkSharing}, which may have an impact on data quality in the long run. 

Some task assignment methods also consider special properties depending on the task. For instance, privacy preservation is an important performance metric for audio transcription tasks~\cite{Celis2016SensitiveTasks}. Others have considered the fairness~\cite{Goel2019CrowdsourcingConstraints}, worker survival or likelihood to continue on tasks~\cite{Kobren2015GettingMoreforLess} and diversity in terms of worker properties~\cite{Barbosa2019RehumanizedLearning}.

\subsection{Worker Performance Modelling}

Based on the complexity and requirements of worker performance estimation method and the task or question assignment method, the literature proposes different ways to represent the quality of each worker, which we summarise below. 

\subsubsection{Worker Probability}

The quality of each worker is modelled by a single attribute that describes the probability of the worker providing the true answer for any given question. This is a simple and widely adopted method~\cite{Guo2012DynamicMaxDiscovery,liu2012CDAS}. However, a single probability score is often insufficient to model the quality of the worker due to variations in question difficulty. The basic worker probability model can be extended by including a confidence value along with the probability value~\cite{Joglekar2015ComprehensiveAlgorithms}. 

Instead of using a single probability value for all the tasks, worker probability can be modelled for each task (\eg~\cite{Mo2013CrosstaskCrowdsourcing}) or question within the task (\eg~\cite{Fan2015}). For example, quality of a specific worker could be 0.5 for sentiment analysis task and 0.8 for classification task.

\subsubsection{Confusion Matrix}

Confusion matrix is extensively used to model worker performance for multiple-choice questions where each question has a fixed number of possible answers (\eg~~\cite{Whitehill2009WhoseExpertise,Raykar2009SupervisedBit,Venanzi2014Community-basedCrowdsourcing}). Each cell $(i,j)$ within the matrix indicates the probability of the worker answering the question with a label $i$ given the true answer of the question is $j$. For the initialisation each worker could be assumed a perfect worker, values could be drawn from a prior distribution, or values can be estimated using gold standard questions.

\subsubsection{Graph-based} 

In a graph-based model, workers or tasks are modelled as nodes in a graph (\eg~\rev{\cite{Cao2012WhomServices,Zhao2014SocialTransfer:Crowdsourcing,Riveni2019}}). Edges represent possible relationships among them. Different approaches are also possible. For instance, task assignments can be modelled as edges in a bipartite graph with both workers and questions as nodes (\eg~\cite{Karger2013EfficientLabeling,Liu2012VariationalCrowdsourcing}).

\subsubsection{Tree-based}

A tree-based model is a slight variant of the graph-based model. For instance, \acite{Mavridis2016UsingCrowdsourcing} uses a skill taxonomy modelled as a tree where nodes represent elementary skills. Each worker also has a set of skills that they possess. A skill distance metric between the required skills for the task and the given skills of a worker is considered as the worker quality value for the particular task.

\subsection{Performance Estimation Methods}
\label{sec:worker-performance-estimation-sub}

Before assigning tasks or questions to workers, we need to estimate the performance of each worker. Estimations can be obtained by using objective measures such as gold standard questions, past/current task performance data, and qualification tests or by using worker characteristics or behavioural traits that are known to correlate with task performance. Table~\ref{tab:worker-estimation} organises prior work based on the performance estimation method.

\begin{table}[h]
    \centering
    \caption{An overview of worker performance estimation methods used in online assignment methods.}
    \label{tab:worker-estimation}
\begin{tabular}{p{5cm}ll}
\toprule
\multicolumn{2}{l}{Method} & Literature\\
\midrule
\multicolumn{2}{l}{Gold Standard Questions \& Qualification Tests}  & \cite{Ipeirotis2014quizz}, \cite{liu2012CDAS} \\
\midrule
Current Answer Distribution &  & \cite{Zheng2015}, \cite{Khan2017CrowdDQS}, \cite{Baba2013}, \cite{raykar2012eliminatingranking} \\
\midrule
\multicolumn{2}{l}{Previous Answers \& Reputation Scores} & \rev{\cite{Peer2014ReputationTurk}, \cite{Schall2012}} \\
\midrule
\multirow{5}{*}{Worker Attributes} & Demographics & \cite{Kazai2012}, \cite{Shaw2011}, \cite{Eickhoff2012}, \cite{Difallah2018Demographics} \\
 & Personality Tests & \cite{Kazai2011}, \cite{Kazai2012}, \cite{Lykourentzou2016} \\
 & Skills & \cite{Mavridis2016UsingCrowdsourcing}, \cite{Kumai2018Skill-and-Stress-AwareStreams} \\
 & Cognitive Tests & \cite{Goncalves2017TaskRouting}, \cite{Hettiachchi2019EffectofCognitive}, \cite{Hettiachchi2020CrowdCog} \\
 & Work Device Features & \cite{Gadiraju2017ModusEnvironments}, \cite{Hettiachchi2020CrowdTasker} \\
 & Worker Context & \cite{Ikeda2017CrowdsourcingPerformance}, \cite{Hettiachchi2020HowWork}\\
\midrule
\multirow{2}{*}{Worker Behaviour} & Behavioural Traces & \cite{Rzeszotarski2011}, \cite{Han2016}, \cite{Gadiraju2019CrowdPre-selection}, \cite{Goyal2018YourAnnotations} \\
 & Social Media Data & \cite{Difallah2013Pick-a-crowd:Networks}, \cite{Zhao2014SocialTransfer:Crowdsourcing}\\
\bottomrule
\end{tabular}
\end{table}

\subsubsection{Gold Standard Questions}

Gold Standard Questions are questions with a known answer. It is common practice to use gold standard questions to estimate worker performance~\cite{liu2012CDAS}. Typically, gold questions are injected into the task to appear among regular questions such that workers are unable to anticipate or detect gold questions. 

When implementing gold standards, it is essential to know how we can inject these questions systematically. Prior work by \acite{liu2013scoring} investigates the optimum number of gold questions to use in a task. It is not beneficial to use a small number of gold standard questions in a large question batch. Workers could then collectively identify and pay more attention to gold questions making them ineffective as quality checks~\cite{checco2018all,Checco2020AdversarialControl}.
Furthermore, creating ground truth data is not straightforward and crowdsourced tasks often do not have ground-truth data. Therefore, scalable and inexpensive methods of creating good gold data are necessary when using gold standards as a quality improvement method. \acite{Oleson2011ProgrammaticCrowdsourcing} present a programmatic approach to generate gold standard data. They report that a programmatic gold method can increase the gold per question ratio, allowing for high-quality data without extended costs.

Instead of populating gold questions before the label collection, we can also validate selected answers using domain experts. For instance, \acite{hung2015minimizing} propose a probabilistic model for classification tasks that help us find a subset of answers to validate through experts. The method considers the output accuracy and detection of inaccurate workers to find the most beneficial answer to validate. In addition, we can use domain experts to generate reliable and high-quality gold data~\cite{hara2013combining}. Finally, in addition to measuring worker performance, gold standard questions can function as training questions that provide feedback to workers ~\cite{Le2010EnsuringDistribution,Gadiraju2015TrainingMicrotasks}.

\subsubsection{Qualification Tests}

Qualification tests contain a set of questions that workers need to complete before accessing the task. A qualification test can contain questions related to worker experience, background or skills that are needed for the actual crowdsourcing task~\cite{Mitra2015}. For instance, a simple language skill test could be an appropriate qualification test for a translation task. A set of gold standard questions can also be presented as a qualification task. As answers are known a-priori, requesters can measure the performance in qualification test and allow a subset of workers to attempt the regular task. Crowdsourcing platforms such as MTurk supports qualification tests.

When using gold standard questions as a qualification test, there should be sufficient coverage of the different questions included in a task. Similarly, the qualification test should be challenging, such that workers are unable to pass is without understanding the task instructions fully.

When employing qualification tests, we can also ask workers to assess their own responses when ground truth data is not available or automated assessment is not feasible. \acite{Gadiraju2017UsingMicrotasks} show that self-assessment can be a useful performance indicator when we account for varying levels of accuracy in worker self-assessments.

\subsubsection{Using Current Answer Distribution}

In an ongoing task, we can also use the current answer distribution to estimate worker accuracy. Expectation Maximisation (EM)~\cite{Dawid1979MaximumAlgorithm} is one of the most commonly used estimation methods to gauge worker performance for multiple class labelling questions (\ie~multiple choice questions)~\cite{Zheng2015}. The method examines all the current answers and iteratively updates worker quality values and task answers until they converge.  \acite{Khan2017CrowdDQS} used a different approach that uses Marginal Likelihood Estimation. They report that compared to Expectation Maximisation, Marginal Likelihood Estimation significantly reduces root mean squared error (RMSE) in predicting worker accuracy when there are few votes per worker. \acite{raykar2012eliminatingranking} considers a discrete optimisation problem and propose a Bayesian approach that can estimate a binary state that decides whether a worker is a spammer or not. 

Estimating worker accuracy from current answer distribution is not exclusive to labelling tasks. \acite{Baba2013} introduced a two-stage workflow with a creation and a review stage for tasks with unstructured responses, such as content generation and language translation. Their method uses the maximum a posteriori (MAP) inference to estimate the accuracy and model parameters.

\subsubsection{\rev{Previous Answers and Reputation Scores}}

\rev{Once crowdsourcing tasks are completed, task requesters can indicate whether they are satisfied with worker responses. Similarly, we can also test if worker responses agree with the majority response. Such signals can be incorporated into a reputation score (\eg~approval rate in Amazon Mechanical Turk platform) and can be used to estimate the worker performance~\cite{Peer2014ReputationTurk}. In addition, research shows that trust relationships among workers can be leveraged as reputation scores~\cite{Schall2012}.}

\subsubsection{Worker Attributes}

When looking at task or question assignment from the workers' perspective, several worker attributes have been shown to have an impact on crowd task performance. 

\begin{itemize}
    \item \textit{Demographics:} \revb{While there is no evidence to support a direct link between demographics and worker performance, literature reports on specific tasks where demographics may influence the performance~\cite{Kazai2012,Shaw2011}}. \revb{Importantly, researchers note that local knowledge~\cite{Goncalves2013CrowdsourcingontheSpot,Goncalves2015MotivatingCrowdsourcing}, language~\cite{Vashistha2017RespeakVoicebasedSpeechTranscription} and work tools~\cite{Gadiraju2017ModusEnvironments} of crowd workers, and the differences in pay rates~\cite{berg2016income,Hosio2014SituatedModel} can lead to location-based performance variations. For example, in a content analysis task that involves assessing US political blogs, \acite{Shaw2011} have shown that US workers unsurprisingly perform significantly better than Indian workers.}
    \revb{In contrast}, in an attempt to examine the preference for games over conventional tasks in relevance labelling, \revb{~\acite{Eickhoff2012} reported no significant difference in the performance of workers from the US} and India in Amazon Mechanical Turk. \revb{Other demographic factors such as age~\cite{Kazai2012} can also influence performance in specific tasks.} Other work have also shown that worker demographics can introduce biases to the data collected~\cite{Difallah2018Demographics,Hettiachchi2019TowardsModeration}.
    \item \textit{Personality:} \acite{Kazai2011} analysed crowd users based on five personality dimensions introduced by Goldberg~\cite{john2008} known as the `Big Five'. They further segmented workers into five types: Spammer, Sloppy, Incompetent, Competent and Diligent based on the personality and reported a significant correlation between the worker type and the mean accuracy of the worker. In a subsequent study, \acite{Kazai2012} also reported that the Big Five personality traits - openness and conscientiousness - are correlated with higher task accuracy. \acite{Lykourentzou2016} also examined the effect of personality on the performance of collaborative crowd work on creative tasks. They created 14 five-person teams: balanced (uniform personality coverage) and imbalanced (excessive leader-type personalities) considering only the outcome of `DISC'~\cite{marston2013} (dominance, inducement, submission, compliance) personality test and reported that balanced teams produce better work in terms of the quality of outcome compared to imbalance teams.
    \item \textit{Cognitive Biases:} The study by \acite{Eickhoff2018} investigates cognitive biases and shows that cognitive biases negatively impact crowd task performance in relevance labelling. Cognitive biases are known as systematic errors in thinking and can impact peoples everyday judgements and decisions.
    \item \textit{Cognitive Ability:} \acite{AlagaraiSampath2014CognitivelyInspiredTask} experiment with task presentation designs relating to cognitive features such as visual saliency of the target fields and working memory requirements. The study conducted on MTurk uses a transcription task and reports  design parameters that can improve task performance.
    \acite{Goncalves2017TaskRouting} investigated the impact of the cognitive ability of crowd worker performance and demonstrated that performance can be predicted from the results of cognitive ability tests. In their study, they used 8 cognitive tests which included visual and fluency tasks and 8 different crowdsourcing task categories attempted by 24 participants in a lab setting. However, they used time-consuming and paper-based cognitive tests from ETS cognitive kit~\cite{Ekstrom1976} that are not practical for an online setting. \acite{Hettiachchi2019EffectofCognitive} investigate the effect of cognitive abilities on crowdsourcing task performance in an online setting. The work leverages the three executive functions of the brain (inhibition control, cognitive flexibility and working memory)~\cite{Diamond2013} to describe and model the relationship between cognitive tests and crowdsourcing tasks. A subsequent study~\cite{Hettiachchi2020CrowdCog} proposes a dynamic task assignment approach that uses cognitive tests. 
    \item \textit{Mood:} Prior work has also investigated if workers' mood has any impact on the crowdsourcing task performance~\cite{Zhuang2019InMengdie}. While there is no evidence that shows a direct link between mood and task accuracy, the study reports that workers in a pleasant mood exhibit higher perceived benefits from completing tasks when compared to workers in an unpleasant mood.
    \item \textit{Work Device Features:} \acite{Gadiraju2017ModusEnvironments} show that crowd work device and its characteristics such as screen size, device speed have an impact on data quality. The research also highlights that the negative impact of bad user interfaces is exacerbated when workers use less suitable work devices. In addition, device sensing capabilities and battery level can also impact the quality of crowd contributions~\cite{Hassani2015Context-AwareCrowdsensing}. \acite{Hettiachchi2020CrowdTasker} explore voice-based crowdsourcing, where workers complete crowd tasks through smart speakers and investigate if there is a performance difference compared to regular crowd work through desktop computers. 
    \item \textit{Worker Context:} Other contextual factors concerning the worker's current situation can also impact crowd task performance. \acite{Ikeda2017CrowdsourcingPerformance} show that task completion rate decreases when workers are busy or with other people. Also, worker context is a critical performance estimator for task assignment in spatial crowdsourcing, where tasks relate to a specific location~\cite{Gummidi2019ACrowdsourcing}. \acite{Hettiachchi2020HowWork} investigate workers' willingness to accept crowd tasks to understand the impact of context when tasks are available through a multitude of work devices. 
    \item \textit{Skills:} Prior work by \acite{Mavridis2016UsingCrowdsourcing} estimates worker performance using a distance measure between the skills of the worker and the skills required for the specific task. They use a taxonomy-based skill model. Similarly, \acite{Kumai2018Skill-and-Stress-AwareStreams} model each skill with a numeric value. For instance, 1 minus the average word error rate (WER) of a worker's typing results can represent their typing skill.
\end{itemize}

\subsubsection{Worker Behaviour}

Prior work shows that worker behaviour data can be used to estimate worker performance~\cite{Gadiraju2019CrowdPre-selection, Rzeszotarski2011, Han2016, Goyal2018YourAnnotations}. \acite{Rzeszotarski2011} proposed `task fingerprinting', a method that builds predictive models of task performance based on user behavioural traces. Their method analyses an array of actions (\eg~scrolling, mouse movements, key-strokes) captured while the user is completing crowdsourcing tasks. Task fingerprinting has been shown to be effective for image tagging, part-of-speech classification, and passage comprehension tasks in Amazon Mechanical Turk. 

\acite{Han2016} also reported that most of the worker behavioural factors are correlated with the output quality in an annotation task. Their method includes several additional features compared to the task fingerprinting method~\cite{Rzeszotarski2011} and uses four types of behavioural features: temporal, page navigation, contextual, and compound. In a different approach, \acite{Kazai2016} show how we can use the behaviours of trained professional workers as gold standard behaviour data to identify workers with poor performance in relevance labelling. 

While other methods~\cite{Han2016,Rzeszotarski2011} aim to classify workers into either `good' or `bad' categories, \acite{Gadiraju2019CrowdPre-selection} classify workers into five categories using behavioural traces from completed HITs. The study shows that significant accuracy improvements can be achieved in image transcription and information finding tasks by selecting workers to tasks based on given categories. To predict task and worker accuracy in relevance labelling tasks, \acite{Goyal2018YourAnnotations} uses action-based (\eg~mouse movement in pixels in horizontal direction, total pixel scroll in vertical direction) and time-based (\eg~fraction of the total time that was spent completing the HIT, mean time between two successive logged click events) features in their predictive model. \acite{Goyal2018YourAnnotations} argue that worker behaviour signals captured in a single session can be used to estimate the work quality when prior work history is unavailable.

Behavioural data like social media interests captured outside the crowdsourcing platforms have also been used to predict task performance~\cite{Difallah2013Pick-a-crowd:Networks}. While this can be an interesting direction which attempts to create a global profile of the crowd worker, current strict privacy regulations would make practical implementation almost impossible.

\subsubsection{Using a combination of estimators}

Rather than using a single performance estimator, it is also possible to use a combination of different estimators. For instance, most of the expectation maximisation based methods use gold standard questions for initial estimation. Similarly, \acite{Barbosa2019RehumanizedLearning} introduces a framework where the worker pool for each task can be constrained using multiple factors such as demographics, experience and skills. Their results show that worker selection with appropriate uniform or skewed populations helps mitigate biases in collected data.

\subsection{Challenges and Limitations \rev{in Performance Estimation}}
\label{sec:challenges-performance-estimation}

While prior work reports promising results on using various worker performance estimation methods, there are many limitations when we consider implementation and broader adoption of such method.

Perhaps the most well-known estimation method is the use of gold standard questions. However, there are several fundamental limitations. First, gold standard questions are not broadly available for all tasks (\eg~tasks with subjective responses). Second, it can be costly to generate good gold questions. Third, gold questions are also susceptible to adversarial attacks. In an attack, workers detect and mark gold standard questions through various third-party tools such that subsequent workers can pay more attention to gold standard questions to amplify their measured performance~\cite{Checco2020AdversarialControl}. 
Despite such limitations, the use of gold standard questions is an effective quality control method applicable to a broader range of tasks.

Worker attributes are also widely used to estimate the worker performance. Attributes like cognitive ability, personality and skills are preferred as they can be extended to estimate task performance across a wider range of tasks. Similarly, task requesters often use demographics (\eg~location, age, education level) as it is straightforward to use them. However, there are notable challenges in integrating certain worker attributes into a task assignment system. For example, attributes like demographics are self-reported by workers, allowing workers to provide incorrect information to gain undue advantages. Comprehensive personality tests are time-consuming and there is also the possibility for workers to manipulate the outcome. Similarly, less competent crowd workers tend to overestimate their performance in self-assessments~\cite{Gadiraju2017UsingMicrotasks}. 

 \rev{In addition, demographics based performance estimation could lead to biased or unfair assignment and discrimination by task requesters leading to fewer tasks of a particular type assigned to workers with specific demographic attributes (\eg~gender, ethnicity, race)~\cite{Hannak2017Bias}. Such unethical approaches and problems should be addressed by crowdsourcing platforms, as well as by researchers.}

Numerous complications exist when concerning the use of worker skills~\cite{Mavridis2016UsingCrowdsourcing,Kumai2018Skill-and-Stress-AwareStreams}. Workers need to list down their skills and such information should be available at platform level. We have to either assume that worker input related to skills are accurate or validate such information. Skill assessment can be a lengthy process increasing the barrier of entry for new workers. Also, requesters have to define which skills are required when creating new tasks.

While worker activity tracking~\cite{Gadiraju2019CrowdPre-selection, Rzeszotarski2011, Han2016, Goyal2018YourAnnotations} has shown promising results, there are several practical limitations. First, such implementations often run as browser-scripts and can make the crowdsourcing platform interface resource intensive. This in turn can limit the accessibility of crowdsourcing platforms, particularly for workers with computing devices with limited capacities and low bandwidth internet connectivity. Second, behavioural data collection, data storage, and performance estimation can be computationally intensive for the back-end infrastructure of the crowdsourcing platforms, thus incurring additional costs. Third, there are privacy concerns with regard to tracking and storing activity data.

\section{Task Assignment Methods}
\label{sec:task-assignment}

In this section, we discuss methods or frameworks that actively prevent contributions of sub-par quality by implementing various quality control mechanisms. In contrast to post-processing techniques, task assignment or routing methods can significantly reduce the overall number of answers required to obtain high-quality output for crowd tasks. Thus, they can bring a financial benefit to task requesters. Also, task assignment can increase the compatibility between worker capabilities and task needs, potentially leading to increased worker satisfaction.

Literature presents a number of task assignment algorithms or frameworks that can be integrated with, or used in place of existing crowdsourcing platforms. They consider different quality metrics (\eg~accuracy, task completion time) and implement one or more quality improvement techniques (\eg~gold standard questions~\cite{Downs2010}, removing or blocking erroneous workers~\cite{Khan2017CrowdDQS}) to enhance the metrics. The primary motivation behind each assignment method can also be divergent. For example, some methods aim to maximise the quality of the output (\eg~\cite{Saberi2017, Fan2015, Zheng2015}) while other methods attempt to reduce the cost by achieving a reasonable accuracy with a minimum number of workers (\eg~\cite{Khan2017CrowdDQS}). 

We organise prior work under task assignment, question assignment and plurality problems we outlined in Section~\ref{sec:problems}. Table~\ref{tab:assignment-methods} provides a brief summary of the worker performance estimation and assignment strategy of each method we discuss in this section.


\begin{center}
{\footnotesize
    \begin{longtable}{cccp{1cm}p{3.2cm}p{3.3cm}p{3.3cm}}
    \caption{An overview of worker performance estimation and assignment strategies of assignment methods.} \label{tab:assignment-methods}\\
    \toprule
        \multicolumn{3}{p{1cm}}{Assignment Problem} & Reference & Performance Estimation & Assignment Strategy & \rev{Method Maturity and Evaluation}  \\
        \parbox[t]{2.5mm}{\rotatebox[origin=c]{90}{Task}} & \parbox[t]{2.5mm}{\rotatebox[origin=c]{90}{Question}} & 
        \parbox[t]{2.5mm}{\rotatebox[origin=c]{90}{Plurality}} \\
        \midrule
        \checkmark&&& \cite{Ho2012OnlineMarkets} & Requesters manually evaluate the answers. &  Based on the online primal-dual framework. & \rev{Basic Research. Offline evaluation using real-world crowdsourcing data.}\\
        \checkmark&&& \cite{Ho2013AdaptiveClassification} & Using gold standard questions. & By extending online primal-dual methods. & \rev{Basic Research. Offline evaluation using simulations and synthetic data.}\\
        \checkmark&&& \cite{Assadi2015OnlineMarkets} & Using bids provided by workers. & Maximises the number of tasks allocated within a budget. & \rev{Basic Research. Offline evaluation using simulations and synthetic data.}\\
        \checkmark&&& \cite{Mo2013CrosstaskCrowdsourcing} & Estimate using the performance in other tasks. & Through a hierarchical Bayesian transfer learning model.  & \rev{Basic Research. Offline evaluation using synthetic and real-world crowdsourcing data.}\\
        \checkmark&&& \cite{Dickerson2018AssigningArrivals} & Use historic records to learn quality distributions. & Model workers and tasks in a bipartite graph and use an adaptive, non-adaptive or greedy method to assign tasks. & \rev{Basic Research. Offline evaluation using real-world crowdsourcing data.}\\
        \checkmark&&& \cite{Hettiachchi2020CrowdCog} & Estimated using cognitive test outcomes. & Select workers to maximise gain in accuracy. & \rev{Prototype Implementation. Online dynamic evaluation with crowd workers.} \\
        \checkmark&&& \cite{Difallah2013Pick-a-crowd:Networks} & Using interested topics captured from social media. & Rank available workers through category-based, expert-profiling and semantic-based assignment models. & \rev{Prototype Implementation. Online dynamic evaluation with crowd workers.}  \\
        \checkmark&&& \cite{Mavridis2016UsingCrowdsourcing} & Through a distance measure between worker skills and the skills required for tasks. & Targets skill compatibility. Assigns specialised tasks to workers with fewer skills first. & \rev{Basic Research. Offline evaluation using synthetic and real-world crowdsourcing data.} \\
        \checkmark&&& \cite{Difallah2016SchedulingHITs} & Assumes that context-switching reduces worker satisfaction and performance. & Scheduling tasks to maximise the likelihood of a worker receiving a task that they have recently worked on.  & \rev{Prototype Implementation. Online dynamic evaluation with crowd workers.} \\
        \checkmark&&& \cite{Difallah2019Deadline-AwareSystems} & Assumes that context-switching reduces worker satisfaction and performance. & Schedule tasks prioritising currently running jobs and workers getting familiar work. & \rev{Prototype Implementation. Online dynamic evaluation with crowd workers.}\\
        \checkmark&&& \cite{Celis2016SensitiveTasks} & Estimate the loss of private information & A graph-based method that maintains privacy without starving the on-demand workforce.  & \rev{Basic Research. Offline evaluation using synthetic and real-world data.} \\
        \checkmark&&& \cite{Kumai2018Skill-and-Stress-AwareStreams} & Estimate worker skills using a qualification task. & Form groups of workers based on skill balance and worker re-assignments. & \rev{Prototype Implementation. Online  evaluation with crowd workers.}\\
        \checkmark&&& \cite{Ikeda2016CollaborativeCrowd4U} & Worker specified task interest and other factors such as skills. & Uses different strategies depending on the task collaboration scheme.  & \rev{Prototype Implementation. No evaluation.} \\ 
        \checkmark&&& \cite{Schmitz2018OnlineCrowdsourcing} & Assumes that expertise of each worker is a known numerical value. & Sequential assignment based on budget, data quality and latency needs. & \rev{Prototype Implementation. Offline evaluation using synthetic and real-world data, and limited online evaluation with crowd workers.}\\
        \midrule
        &\checkmark&\checkmark&\cite{liu2012CDAS} CDAS & Injecting gold standard questions. & Estimate the required answer count and use early termination. & \rev{Prototype Implementation. Online dynamic evaluation with crowd workers.}\\
        &\checkmark&&\cite{Khan2017CrowdDQS} CrowdDQS & Marginal likelihood curve estimation. & Maximise gain in accuracy.  & \rev{Prototype Implementation. Online dynamic evaluation with crowd workers.}\\
        &\checkmark&&\cite{Fan2015} iCrowd  & Static gold standard questions \& task similarity. & Save questions for most accurate workers.  & \rev{Prototype Implementation. Online dynamic evaluation with crowd workers.}\\
        &\checkmark&&\cite{Saberi2017} OSQC& Hybrid gold plurality algorithm & Multi-rule quality control.  & \rev{Basic Research. Offline evaluation using real-world crowdsourcing data.}\\
        &\checkmark&&\cite{Chen2013OKG} OKG & Statistical inference with Beta distribution priors. & Maximise gain in accuracy. & \rev{Basic Research. Offline evaluation using synthetic and real-world crowdsourcing data.}\\
        &\checkmark&&\cite{Zheng2015} QASCA & Expectation maximisation (EM). & Maximise gain in accuracy or F-score.  & \rev{Prototype Implementation. Online dynamic evaluation with crowd workers.}\\
        &\checkmark&&\cite{Ipeirotis2014quizz} Quizz & Estimate using only gold standard question responses. & Maximise information entrophy. & \rev{Prototype Implementation. Online dynamic evaluation with crowd workers.} \\
        &\checkmark&&\cite{Goel2019CrowdsourcingConstraints} & Estimate using limited gold standard questions. & Maximise gain in accuracy while satisfying budget, fairness and diversity constraints. & \rev{Basic Research. Offline evaluation using real-world crowdsourcing data.}\\
        \midrule
        &&\checkmark& \cite{Mo2013Optimizing} & Using gold standard questions. & Estimate plurality form a greedy algorithm that assumes that answer quality increases monotonically at a decreasing rate with its plurality. & \rev{Basic Research. Offline evaluation using synthetic and real-world crowdsourcing data.} \\
        &&\checkmark & \cite{Siddharthan2016CrowdsourcingSize} & By modelling task difficulty and worker skills. & Through an incremental Bayesian model that re-evaluate answer quality at each stage.  & \rev{Basic Research. Offline evaluation using real-world crowdsourcing data.}\\
        &&\checkmark& \cite{Tu2019Quality-AssuredCrowdsourcing} & By iteratively estimating worker expertise and question difficulty. & Batch assignment maximising the number of questions completed in each batch.  & \rev{Basic Research. Offline evaluation using synthetic and real-world crowdsourcing data.} \\
        &&\checkmark& \cite{Abraham2016} & Assumes the past performance of a worker is known. & Decide on when to stop assigning another worker. & \rev{Basic Research. Offline evaluation using synthetic and real-world crowdsourcing data.}\\
        \bottomrule
    \end{longtable}}
\end{center}

\subsection{Heterogeneous Task Assignment}

As crowdsourcing platforms contain a variety of tasks (\eg~sentiment analysis, classification, transcription), heterogeneous task assignment focuses on matching different task types with workers. Heterogeneous task assignment can be particularly useful in cases where `expert' workers must be allocated for more difficult tasks~\cite{Ho2012OnlineMarkets}. In addition to heterogeneous task assignment, crowdsourcing literature also explores question assignment, where questions within the same task (\eg~different questions of sentiment analysis task) are assigned to different workers to maximise the performance gain. We also review question assignment methods in Section~\ref{section:question-assignment}.

Task assignment involves multiple steps. First, worker performance is modelled and estimated using different methods discussed in Section~\ref{sec:worker-performance-estimation}. Then, the task assignment process is carried to maximise the potential gain in terms of a specific performance criteria. For instance, one task assignment method could achieve modest data quality gains while minimising the overall cost. In contrast, another method could aim to achieve the highest possible data quality with a set budget.

\acite{Ho2012OnlineMarkets} propose a task assignment method based on the online primal-dual framework, which has been previously utilised for different online optimisation problems. The proposed Dual Task Assigner algorithm assumes that workers with unknown skills request tasks one at a time. In the study, researchers use three types of ellipse classification tasks to account for different expertise levels and use a translation task to simulate different skills. However, their approach assumes that the requester can immediately evaluate the quality of completed work. This vastly limits the applicability of their approach in a real-world crowdsourcing problem. \acite{Ho2013AdaptiveClassification} further investigate heterogeneous task assignment in classification tasks with binary labels. However, for the assignment, they use gold standard questions of each task type to estimate the accuracy of the workers.

We can also examine task assignment from the requester perspective. \acite{Assadi2015OnlineMarkets} propose an online algorithm that can be used by a requester to maximise the number of tasks allocated with a fixed budget. In a different approach for task assignment, \acite{Mo2013CrosstaskCrowdsourcing} apply a hierarchical Bayesian transfer learning model. They use the historical performance of workers in a similar or different type of tasks to estimate the accuracy of the new tasks. Their experiment with a real-world dataset shows the effectiveness of the proposed approach when transferring knowledge from related but different crowd tasks (\eg~questions on sports vs makeup and cooking). However, their real-world evaluation is limited to a single scenario with one source task and one target task. 

While most methods focus on a predefined set of tasks, \acite{Dickerson2018AssigningArrivals} examine task assignment when tasks are not known a-priori. Their work proposes a novel theoretical model, called Online Task Assignment with Two-Sided Arrival (OTA-TSA), where both workers and tasks arrive in an online manner.

Data collected outside crowdsourcing platforms can also be used to match tasks with workers. \acite{Difallah2013Pick-a-crowd:Networks} present a system where tasks are allocated based on worker profile data such as interested topics captured from a social media network. Similarly, \acite{Zhao2014SocialTransfer:Crowdsourcing} propose `Social Transfer graph' for task matching. They demonstrate how tasks on Quora can be matched with Quora users' by extracting respective users' Twitter profile data (\ie~tweets and connections). The general applicability of such methods raises numerous practical and ethical considerations. 

\acite{Mavridis2016UsingCrowdsourcing} introduced a skill-based task assignment model. Worker performance is estimated using a distance measure between the skills of the worker and the skills required for the specific tasks. The method attempts to assign the most specialised task first to the workers with the lowest number of skills based on the distance measure.

Task assignment can be challenging for more complex and collaborative tasks. \acite{Ikeda2016CollaborativeCrowd4U} propose a task assignment framework that can decompose complex tasks and support sequential, simultaneous and hybrid worker collaboration schemes. Their assignment strategy selects a worker based on interests indicated by workers and their eligibility calculated using the project description and worker human factors (\eg~language skills). In contrast, \acite{Schmitz2018OnlineCrowdsourcing} look at non-decomposable macro-tasks like document drafting. They propose a sequential assignment model, where multiple workers attempt a task on a fixed time-slot, one after the other. At the end of each iteration, the next worker is selected if the task does not meet the desired quality threshold.

Instead of assigning tasks on the fly, it is also possible to schedule them when tasks are known apriori. Prior work by \acite{Difallah2016SchedulingHITs} investigates task scheduling in crowdsourcing platforms and shows that scheduling can help minimise the overall task latency, while significantly improving the worker productivity captured through average task execution time. Research also highlights that scheduling is useful in ensuring tasks are fairly distributed across workers~\cite{Difallah2019Deadline-AwareSystems}. 

Addressing the growing concerns on crowdsourcing sensitive tasks like transcribing audio scripts, \acite{Celis2016SensitiveTasks} examined task assignment with regard to trade-off in privacy. To preserve content privacy, we need to ensure that not too many parts of the same job are assigned to the same worker. They introduced three settings: PUSH, PULL, and a new setting, Tug Of War (TOW), which aims to balance the benefit for both workers (by ensuring they can attempt a reasonable number of questions) and requesters (by minimising the privacy loss).

Instead of assigning tasks to individual workers, \acite{Kumai2018Skill-and-Stress-AwareStreams} investigate the worker group assignment problem, where task requesters should select a group of workers for each task. They represent the worker accuracy using skills estimated through a qualification task and then forms groups based on three strategies that consider the skill balance among groups and the number of worker re-assignments.

\subsection{Question Assignment}
\label{section:question-assignment}

The aim of question assignment is to match workers with questions within a task such that we can obtain high-quality output. Unlike in heterogeneous task assignment, we need to estimate worker performance and allocate tasks as workers complete submit answers to individual or batches of questions. \acite{Zheng2015} present a formal definition of question assignment problem in crowdsourcing and show that optimal question assignment is an NP-hard problem. 

Question assignment involves several fundamental steps. First, we obtain a set of questions that are available to be assigned. Such candidate questions should not have been previously assigned to the current worker and should have available assignments with respect to the maximum number of answers required. Second, we estimate the performance gain (in terms of accuracy, for example) for each candidate question. Third, we select a subset of questions to be assigned to the given workers.

Baseline approaches for question assignment are random assignment or a round robin assignment. Typical crowdsourcing platforms use these baseline approaches for question assignment.  

\subsubsection{Assigning questions to workers in a sequential manner}

The question assignment problem can vary depending on the worker arrival assumption. The most practical problem is how to find a suitable question or a specific number of questions for an individual worker given a set of candidate questions. A naive way to assign questions is to enumerate all feasible assignments, calculate the performance gain for each assignment and then picks the assignment with the maximum performance gain. However, this method is computationally expensive and is not practical for typical crowdsourcing platforms where each task has a large number of questions. 

\acite{Zheng2015} proposed a question assignment framework (QASCA) which attempt to maximise either accuracy or F-score. For assigning $k$ questions based on accuracy, the paper proposes the Top-K benefit algorithm which calculates the gain in expected number of correct answers for each question in candidate set and pick the questions which have the highest benefits. The algorithm has a time-complexity of $O(n)$ where n is the number of questions in the candidate set. A more complex online algorithm is presented for assigning questions based on F-score. 

`CrowdDQS' proposed by \acite{Khan2017CrowdDQS} is a dynamic question assignment mechanism which examines most recent votes and selectively assigns gold standard questions to workers to identify and removes workers with poor performance in real-time . They claim the proposed system which integrates seamlessly with Mechanical Turk can drastically reduce (up to 6 times) the number of votes required to accurately answer questions when compared to a round-robin assignment with majority voting. The proposed question assignment method aims to maximise the potential gain. The algorithm greedily chooses a question from the candidate set whose confidence score stands to increase the most if another answer is obtained from the considered worker.

Another dynamic question assignment method proposed by \acite{Kobren2015GettingMoreforLess} uses the worker survival metric (a user’s likelihood of continuing to work on a task). Survival score is formulated using different measures such as accuracy, response time, the difficulty of recently completed questions. The framework assigns questions to workers in order to achieve higher worker engagement and higher value for the task requester. Modelled using the markov decision process, the method aims to assign a question that maximises worker survival and expected information gain.

Different questions within a task may require knowledge and expertise on various domains. The task assignment method by \acite{Zheng2016DOCS:Bases} attempts to organise questions and workers into different domains by building a knowledge base. Questions with uncertain true labels are then assigned to workers when their expertise overlap with the question's domain.

\subsubsection{Question Assignment with a batch of workers}

Another variant of the question assignment problem is to come up with an optimal assignment scheme given a set of workers and set of questions as opposed to assigning for a sequence of workers (\eg~\cite{Zhuang2015,Khan2017CrowdDQS}). \acite{Cao2012WhomServices} termed this as the Jury Selection Problem (JSP) where they aim to select a subset of crowd workers for each question under a limited budget, whose majority voting aggregated answers have the lowest probability of producing an incorrect answer.

\acite{Fan2015} introduced dynamic crowdsourcing framework named `iCrowd' which assigns tasks to workers with a higher chance of accurately completing the task using a graph based estimation model. They consider the task similarity when estimating worker accuracy. The proposed question assignment strategy has three steps. First, it identifies a set of active workers who are ready to work on the task and dynamically finds sets of workers with the highest estimated accuracy for each available question. Then, the framework uses a greedy-approximation algorithm to formulate the optimum assignments ensuring each worker has no more than one question. Then, it strategically assigns gold standard questions to workers who are left without any question assignments.

`AskIt' proposed by \acite{Boim2012AskIt} is another framework that achieves batch-wise question assignment. The assignment method aims to minimise the global uncertainty of entropy for questions while satisfying general assignment constraints such as maximum number of answers required for each question. Two metrics are proposed to measure global uncertainty that uses the difference between maximum and minimum entropy for individual questions. AskIt uses a greedy-heuristic to come up with the optimum assignment scheme. In addition, the framework employs an initial pre-processing step that uses collaborative filtering to predict missing answers and to identify questions that are likely to be skipped by a specific worker. However, we note that the paper lacks details of the question assignment algorithm.

\acite{Goel2019CrowdsourcingConstraints} proposed an algorithm for assigning tasks to workers, that optimises the expected answer accuracy while ensuring that the collected answers satisfy pre-specified notions of error fairness. The algorithm also limits the probability of assigning many tasks to a single worker, thus ensuring the diversity of responses. Question assignment is modelled as a constrained optimisation problem that finds the optimal crowdsourcing policy.

In a different approach, the method proposed by \acite{Li2014} assigns a portion of questions to the entire worker pool and estimates the accuracy for sub-groups of workers based on characteristics such as nationality, education level and gender. Then, the framework assigns questions to workers from the specific sub-group with the highest information gain. However, this method is not practical and cost effective when considering implementation on a crowdsourcing platform with a large number of workers from diverse backgrounds~\cite{Difallah2018Demographics}.

\subsubsection{Blocking or Removing workers}

Question assignment can also be achieved by blocking or removing workers from the pool of eligible workers as opposed to actively assigning questions to workers. CrowdDQS~\cite{Khan2017CrowdDQS} uses this blocking technique to further improve assignment performance. 
\acite{Saberi2017} proposed a statistical quality control framework (OSQC) for multi-label classification tasks which monitors the performance of workers and removes workers with high error estimates at the end of processing each batch. They propose a novel method to estimate the worker accuracy -- the hybrid gold plurality algorithm which uses gold standard questions and plurality answer agreement mechanism. Question assignment is based on a Multi-rule Quality Control System which assigns a value (0,1) to the worker at the end of each batch based on the past error rate and the estimated current error rate. Early termination is also another similar strategy where workers can no longer provide answers to a particular question which already has an answer with sufficient certainty~\cite{liu2012CDAS}.


\subsubsection{Question Assignment with Budget Constraints}

\acite{Qiu2016CrowdSelect:Selection} investigate binary labelling tasks. Their proposed method uses previously completed gold standard questions and estimated labels from task requesters to calculate the historic error rate for workers. Then, predicts worker error rate for upcoming questions, through an auto-regressive moving average (ARMA) model. Questions are assigned by maximising the accuracy with respect to the limited budget when worker payment is not constant.

\acite{Rangi2018Multi-armedAbility} approach task assignment with a multi-arm-bandit setup and propose using the simplified bounded KUBE (B-KUBE) algorithm as a solution. In their method, workers indicate their interest in doing the tasks, quote their charges per task, and specify the maximum number of questions they are willing to answer. Worker accuracy is estimated using the current answer distribution.

Similarly, \acite{Singer2013PricingMechanisms} propose a pricing framework when workers bid for tasks with their expected reward and the number of questions they wish to uptake. Their method aims to maximise the number of questions completed under a fixed-budget or minimise payments for a given number of tasks. 

\subsubsection{Assigning Gold Standard Questions}

\rev{Instead of assigning individual questions, we can also assign a specific type of question.} Some frameworks have investigated whether to assign a golden standard question or a regular question when a worker requests a task. \acite{Ipeirotis2014quizz} presented `Quizz', a gamified crowdsourcing system for answering multiple choice questions. The framework uses a Markov Decision Process to select the next action. \rev{However, gold standard based question assignment alone may not lead to improved data quality due to inherent limitations in gold standard questions discussed in Section~\ref{sec:challenges-performance-estimation}.}

\subsubsection{Other Approaches}

\acite{Kang2017SequentialMulticlassLabeling} introduce a game based sequential questioning strategy for question assignment in multi-class labelling questions. They convert the questions into a series of binary questions and demonstrate the reliability of their proposed approach which considers worker responses at each step.

\subsection{Plurality Assignment}

Crowd task accuracy can be improved by obtaining multiple answers from different workers for the same question. In a typical crowdsourcing platform, the number of answers required for each task is set by task requester prior to the task deployment. However, due to variations in worker capabilities and question difficulty~\cite{Tu2019Quality-AssuredCrowdsourcing}, some questions may require more answers, whereas few answers would be sufficient for the others. Crowdsourcing research that addresses the plurality assignment problem~\cite{Mo2013Optimizing} aim to dynamically decide how many answers are needed for each question. 

For binary labelling tasks, \acite{liu2012CDAS} estimate the number of answers required for each question before conducting question assignment. They introduce two prediction models (basic model and an optimised version) that use workers' accuracy distribution. As such accuracy distributions are generally not available in crowdsourcing platforms, a sampling method is used to collect the accuracy of available workers.

\acite{Mo2013Optimizing} propose a dynamic programming based approach to address the plurality assignment problem while maximising the output quality under a given budget. The paper identifies two key properties in crowdsourcing tasks, monotonicity and diminishing returns that describe a question with the final answer quality increasing monotonically at a decreasing rate with its plurality. They also propose an efficient greedy algorithm that can provide near optimal solutions to plurality assignment problem when monotonicity and diminishing returns properties are satisfied. 

Similarly, \acite{Siddharthan2016CrowdsourcingSize} presents an incremental Bayesian model that estimates the plurality for a classification task with a large number of categories. Results obtained through their method outperforms majority voting and is comparable to a different Bayesian approach (\ie~standard multinomial naive Bayes (MNB)) that uses a larger fixed answer count. 

Worker expertise and question difficulty are two key variables that impact the confidence of an answer and plurality. In a batch-processing approach, prior work by \acite{Tu2019Quality-AssuredCrowdsourcing} efficiently estimated these two parameters to maximise the number of questions reliably answered at the end of each batch. \rev{In each batch, the proposed dual-cycle estimation method iteratively estimates inference between worker expertise and answer confidence, and the inference between question easiness and answer confidence in two separate cycles.}

Instead of determining the plurality before task deployment, we can dynamically decide and limit the number of answers that we collect for each question. \acite{Abraham2016} proposed an adaptive method that considers the differences and uncertainty of the answers provided and decide on when to stop assigning another worker for the task.

\subsection{Challenges and Limitations \rev{in Task Assignment}}

We discuss general challenges and limitation in task assignment methods. There is no straightforward, low-cost and effective solution for task assignment~\cite{Fan2015}. Therefore, each method and evaluation has their merits and limitations. 

Concerning worker accuracy estimation, some studies infer worker quality instead of objectively estimating them. For example, \acite{Saberi2017} evaluate their statistical quality control framework proposed with crowd workers on Mechanical Turk where they simulate the past error rates of workers who completed the task using a standard normal distribution. Similarly, prior work by \acite{Schmitz2018OnlineCrowdsourcing} treats the work quality assessment step as a black-box process and assumes the expertise of each worker as a known numerical value. In both cases, it is difficult to argue that findings of such studies hold in real-word crowd platforms due to broader variations in crowd worker quality.

Some studies (\eg~\cite{Boim2012AskIt,Ho2013AdaptiveClassification,Assadi2015OnlineMarkets}) evaluate task assignment methods using synthetic data instead of using a real-time deployment or a crowdsourced dataset. 
Furthermore, as popular crowdsourcing platforms including Amazon Mechanical Turk do not provide sufficient means to dynamically assign tasks, all the aforementioned studies (\eg~\cite{Khan2017CrowdDQS,Zheng2015,Fan2015}) have evaluated their proposed frameworks using the external question feature of these platforms. While this is the standard for crowdsourcing research, it is unclear how worker behaviour in controlled studies compares with regular task performance.

While certain assignment methods (\eg~\cite{Khan2017CrowdDQS}) use random or fixed values for the initial worker accuracy, other methods (\eg~\cite{Fan2015,Ipeirotis2014quizz}) use gold standard questions. Gold standard questions are widely used in crowdsourcing platforms. However, as discussed in Section~\ref{sec:worker-performance-estimation}, there are inherent limitations that make the use of gold questions less desirable. Also, some other methods use historic records~\cite{liu2012CDAS} and suffer from the cold-start problem. These methods do not work with new workers in a crowdsourcing platform.

\subsubsection{Heterogeneous Task Assignment Challenges}

Different worker performance estimation strategies (\eg transfer learning from similar tasks~\cite{Mo2013CrosstaskCrowdsourcing}, worker attributes~\cite{Mavridis2016UsingCrowdsourcing,Hettiachchi2020CrowdCog}) are useful for task assignment. Literature only shows that they can work on specific task types. For example, real world evaluation by \acite{Mo2013CrosstaskCrowdsourcing} is limited to a single source and target task pair.

Overall, heterogeneous task assignment is a highly desirable approach that can potentially work across a broader range of tasks. However, more evidence and experiments are needed to show that they work with various tasks (\eg~Prior work by \acite{Hettiachchi2020CrowdCog} uses four types of common crowdsourcing tasks) and can sustain performance over time.

\subsubsection{Question Assignment Challenges}

Question assignment methods continuously monitor worker answers and create assignments at each step, making them typically more effective than heterogeneous task assignment methods. However, key challenges in adopting question assignment are the complexity in implementation and the cost of calculating the assignments. For example, even with an efficient question assignment algorithm solution such as QASCA~\cite{Zheng2015}, assignment time linearly increase with the number of questions. Therefore, computational complexity is an important factor to consider when employing question assignment methods in a real world system. 

The majority of question assignment methods are also limited to multi-class labelling problems~\cite{Zheng2015,Khan2017CrowdDQS,Ipeirotis2014quizz, Kobren2015GettingMoreforLess}. While literature argues that other types of tasks (\eg~a continuous value) can be converted to multi-class or binary labelling problems~\cite{Zheng2015}, there is no research that shows that question assignment methods can work in such cases.


\subsubsection{Plurality Assignment Challenges}

Plurality assignment is an important problem in crowdsourcing. Proposed methods aim to estimate plurality either upfront~\cite{liu2012CDAS} or during the task execution~\cite{Abraham2016,Tu2019Quality-AssuredCrowdsourcing} which can help reduce the overall cost for task requesters. Similar to question assignment, estimating plurality is often investigated considering multi-class labelling questions. While it is feasible to estimate plurality for labelling questions, it is far more complicated for crowd tasks that involve complex inputs, such as audio tagging and semantic segmentation. However, plurality assignment solutions are also more valuable for such tasks as each response involves a higher work time and reward. 

As plurality assignment solutions do not achieve specific worker-question match, they are less complicated than question assignment methods. Plurality assignment solutions can also be more effective when implemented together with question or task assignment methods~\cite{liu2012CDAS}. However, further research is needed to ensure their utility in a dynamic online setting.

\section{Crowdsourcing Platforms}
\label{sec:platforms}

In this section, we briefly review existing crowdsourcing platforms and standard task assignment mechanisms available in them. At a high level, current crowdsourcing platforms do not support complex task assignment methods proposed in the literature. However, certain functionalities and limited assignment methods are available to task requesters.

In Amazon Mechanical Turk\footnote{https://www.mturk.com/}, requesters can use task pre-qualifications to limit the workers who are able to see and attempt their task. The platform provides a set of pre-specified qualifications such as worker historical approval rate, location and sex. In addition, task requesters can create custom qualifications and include workers based on previous tasks or qualification tests. Further, by using MTurk API and other third-party libraries and tools (\eg~PsiTurk~\cite{Gureckis2016PsiTurk:Online}), task requesters can build advanced task assignment methods on top of MTurk.

Toloka by Yandex\footnote{Launched by Yandex in 2014. https://toloka.ai/} is another popular crowdsourcing platform. Toloka allows task requesters to set-up worker skills that gets automatically updated based on 
the rate of correct responses (with gold standard questions, majority vote, or post-verification) and behavioural features like fast responses. Requesters can also configure rules based on skills. For example, rules could automatically block workers from the task if their skill level drops below a given threshold\footnote{https://toloka.ai/crowdscience/quality}. In addition, Toloka also provides a feature called `incremental relabeling' to facilitate dynamic plurality.

\rev{Microworkers\footnote{https://www.microworkers.com/} is a similar crowdsourcing platform that provides a large collection of task templates. To facilitate basic task assignment, the platform allows custom worker groups, where requesters direct new tasks to workers who have provided satisfactory output in previous tasks.} Prolific\footnote{https://www.prolific.co/} is another crowdsourcing platform that is tailored for surveys and research activities. The platform provides more than 100 demographic screeners to ensure the task is assigned for a restricted worker pool. 


Other commercial crowdsourcing platforms such as Scale\footnote{https://scale.com/}, Appen\footnote{Previously Figure Eight and CrowdFlower. https://appen.com/} and Lionbridge AI\footnote{https://lionbridge.ai/} focus on providing an end-to-end service to task requesters. They use a combination of crowdsourced and automated approaches to complete the task. While implementation details are not available, such platforms also utilise task assignment strategies where they use automated approaches for simpler elements of the work pipeline and get crowd workers to attempt difficult parts such as quality control, edge cases, and complex data types\footnote{https://scale.com/blog/scaling-menu-transcription-tasks-with-scale-document}.

Further, in crowdsourcing platforms that focus on complex tasks and projects (\eg~Upwork, Freelancer, Fiverr), task assignment is explicit. Task requesters examine the candidate workers who express willingness to complete the task and assign the task to one or more workers based on their profile. This manual assignment process is only practical for complex tasks that involve specialised workers, longer task times and higher rewards. \rev{Table~\ref{tab:platforms} summarises task assignment methods offered in current commercial crowdsourcing platforms.}

\begin{table}[htb]
    \centering
    \caption{\rev{Task assignment capabilities available in existing commercial platforms.}}
    \label{tab:platforms}
    \begin{tabular}{ll}
    \toprule
        \rev{Platform} & \rev{Task Assignment Methods Available}\\
        \midrule
        \rev{Amazon Mechanical Turk}  & \rev{Task pre-qualifications, Third-party integrations using the API}\\
        Appen (previously & Gold standard questions\\
        Figure 8, CrowdFlower)  & \\
        \rev{Microworkers} & \rev{Assign to a custom worker group} \\
        \rev{Prolific} & \rev{Demographic screeners} \\
        \rev{Toloka by Yandex} & \rev{Worker skill based task assignment, Gold standard questions} \\
        \bottomrule
    \end{tabular}
\end{table}


\section{Future Directions}
\label{sec:future}

When discussing the future of crowd work, \acite{Kittur2013TheWork} identify task assignment as one of the key elements that can improve the value and meaning of crowd work. While task assignment has been increasingly researched in recent years, we do not see widespread adoption of task assignment strategies in commercial crowdsourcing platforms~\cite{Daniel2018}. In this section, we reflect on limitations with current approaches and discuss how future research could address them to promote the practical use of task assignment.

One of the critical limitations of many task assignment methods is that they fail to work across a broader range of tasks. Thus, there is little incentive for crowdsourcing platforms to implement or facilitate such methods. Future work could explore more generalisable methods that do not directly depend on the task (\eg~cognitive test based task assignment~\cite{Hettiachchi2020CrowdCog}). Research should also focus on how to address the cold start issue in crowdsourcing task assignment. Particularly, task requesters often do not have the luxury of collecting large volumes of training data or accessing and analysing past worker records before employing a task assignment method. Therefore, new methods that work with generic models would be more favourable to requesters.

Moreover, integrating different worker accuracy estimation methods and task assignment strategies is another feasible research direction that can further improve the value and utility of assignment methods. For example,\acite{Barbosa2019RehumanizedLearning} attempt to integrate worker demographics and related attributes and show that we can improve data quality by allowing requesters to pre-specify the workforce diversity or uniformity. Similarly, research shows how cognitive~\cite{Hettiachchi2020CrowdCog}, personality~\cite{Kazai2012}, and task-specific qualification tests~\cite{Mitra2015} are good indicators of worker performance. Future work could investigate how to encapsulate different test scores to provide a unified estimation of worker accuracy. A prudent strategy is to implement a test marketplace, where task requesters could publish different tests that other requesters can use. 

While crowdsourcing is an effective method to harness large volumes of training data for machine learning models~\cite{vaughan2017making}, different biases (\eg~population bias, presentation bias) can be introduced through crowdsourced data collection process~\cite{Olteanu2019SocialBoundaries,Mehrabi2019ALearning}. While biases can be identified~\cite{Hu2020CrowdsourcingDatasets} and reduced in post-processing steps such as aggregation~\cite{Kamar2015IdentifyingCrowdsourcing}, future research should explore how task assignment methods can proactively manage such biases~\cite{Goel2019CrowdsourcingConstraints}.

Furthermore, due to limited features and the competitive nature in crowdsourcing platforms, workers tend to use numerous third-party tools to increase their productivity~\cite{Kaplan2018StrivingWorkers}, leading to task switching behaviour and increased fragmentation in work-life balance~\cite{Williams2019TheCrowdwork}.
It is important to consider worker factors, and develop approaches that can potentially help workers manage their work (\eg~task scheduling approaches that help reduce context switching~\cite{Difallah2016SchedulingHITs}, flexible ways of conducting crowd work~\cite{Hettiachchi2020CrowdTasker}).

Finally, fair compensation for crowd workers is another important aspect~\cite{whiting2019fair,salehi2015wedynamo}. However, it is not sufficient to ensure that worker earnings meet the minimum hourly pay rate, requesters and platforms need to help them minimise the idle time in between jobs. In fact, task assignment reduces task search time by matching workers to compatible tasks. Future work could explore and quantify how such factors are improved through task assignment. Furthermore, assignment methods should explore task matching at a more granular level~\cite{Gadiraju2019CrowdPre-selection,Hettiachchi2020CrowdCog,Khan2017CrowdDQS} than simply identifying `good' or `bad' workers~\cite{Rzeszotarski2011}. This will be particularly beneficial for inexperienced workers as well as others who may not be universally good at all tasks.

\section{Conclusion}
\label{sec:conclusion}

Data quality improvement methods are employed at different stages of the crowdsourcing life cycle. In this review, we provide an extensive overview of online task assignment methods in crowdsourcing that are employed during task deployment. Starting with a succinct overview of data quality improvement methods in crowdsourcing, we dissect online methods into heterogeneous task assignment, question assignment and plurality assignment problems. We discuss the challenges and limitations of existing task assignment methods, particularly their applicability, complexity, effectiveness, and cost. We anticipate that our review and discussions will help researchers and practitioners understand and adopt specific assignment methods to work for their needs. Finally, we detail a set of future research directions in crowdsourcing task assignment highlighting how research can further establish that task assignment methods are broadly applicable, beneficial to workers, and capable of mitigating biases in data.

\bibliographystyle{ACM-Reference-Format}
\bibliography{bibliography}

\end{document}